\documentclass[11pt,english,super, sort&compress]{article}

\usepackage{graphicx}
\usepackage{amssymb,amsmath}
\usepackage{bbm}
\usepackage{float}
\usepackage{bm}
\usepackage{setspace}
\usepackage{parskip}
\usepackage{booktabs}

\usepackage{xcolor}
\definecolor{midnightblue}{cmyk}{1,1,0,0.1}
\definecolor{forestgreen}{cmyk}{0.76,0,0.26,0.5}

\usepackage[a4paper]{geometry}
\usepackage{epsfig}
\usepackage[numbers]{natbib}
\usepackage{lineno}
\usepackage{hyperref}
\hypersetup{
    bookmarks=true,         
    unicode=false,          
    pdftoolbar=true,        
    pdfmenubar=true,        
    pdffitwindow=false,     
    pdfstartview={FitH},    
    pdftitle={My title},    
    pdfauthor={Author},     
    pdfsubject={Subject},   
    pdfcreator={Creator},   
    pdfproducer={Producer}, 
    pdfkeywords={keyword1} {key2} {key3}, 
    pdfnewwindow=true,      
    colorlinks=true,       
    linkcolor=midnightblue,          
    citecolor=magenta,        
    filecolor=midnightblue,      
    urlcolor=midnightblue,          
}

\usepackage{listings}
\definecolor{mygreen}{rgb}{0,0.6,0}
\definecolor{mygray}{rgb}{0.5,0.5,0.5}
\definecolor{mymauve}{rgb}{0.58,0,0.82}
\makeatletter
\usepackage{lineno}
\makeatother
\usepackage{babel}

\usepackage{mathabx}

\begin{document}

\par
\begin{singlespace}
\begin{center}
{\LARGE Deep Potential: a general representation of a many-body potential energy surface}

\vspace{0.2 in}
Jiequn Han$^{1,*}$, Linfeng Zhang$^{1,*}$, Roberto Car$^{1,2}$, and Weinan E$^{1,3,4,\dagger}$\\
\vspace{0.2 in}
 $^{1}$ Program in Applied and Computational Mathematics, \\Princeton University, Princeton, NJ 08544, USA.\\
 $^{2}$ Department of Chemistry, Department of Physics, and Princeton Institute for the Science and Technology of Materials, Princeton University, Princeton, NJ 08544, USA\\
 $^{3}$ Department of Mathematics, Princeton University, Princeton, NJ 08544, USA\\
 $^{4}$ Center for Data Science and Beijing International Center for Mathematical
Research, Peking University,  and Beijing Institute of Big Data Research, 
Beijing, 100871, China\\
$^*$ These authors contributed equally to this work.\\
 $^\dagger$ Email: \texttt{weinan@math.princeton.edu}
\\

\end{center}
\end{singlespace}
\bigskip
\doublespacing

\begin{center} 
ABSTRACT 
\end{center}
We present a simple, yet general, 
end-to-end deep neural network representation of
the potential energy surface for atomic and molecular systems. 
This methodology, which we call \textit{Deep Potential}, 
is ``first-principle" based, in the sense that no \textit{ad hoc}
approximations or empirical fitting functions
are required. The neural network structure naturally respects the
underlying symmetries of the systems. 
When tested on a wide variety of examples, 
Deep Potential is able to reproduce the original
model, either empirical or quantum mechanics based,  
within chemical accuracy.
The computational cost of this new model 
is not substantially larger than that of empirical force fields. 
In addition, the method has promising scalability properties.
This brings us one step closer to being able to carry out molecular simulations
with accuracy comparable to that of quantum mechanics models and 
computational cost comparable to that of empirical potentials.

\section{\label{sec1}Introduction}
A representation of the potential energy surface (PES) for general systems of atoms or molecules 
is the basic building block for molecular dynamics (MD) and/or Monte Carlo (MC) simulations, 
which are common tools in many disciplines, including physics, chemistry, biology, and materials science. 
Until now, this problem has been addressed using two very different approaches.  
At one extreme, 
empirical potentials have been constructed by fitting limited experimental and/or numerical data from accurate quantum mechanical calculations. 
Well-known examples include the Lennard-Jones potential\cite{jones1924LJ}, 
the Stillinger-Weber potential\cite{stillinger1985SW}, the embedded-atom method (EAM) potential\cite{daw1984EAM}, 
the CHARMM\cite{mackerell1998charmm}/AMBER\cite{wang2000amber} force fields, 
the reactive force fields\cite{van2001rff}, etc. 
These potentials are numerically efficient, 
allowing large-scale simulations (up to millions of atoms), 
but their construction is very much an art 
and their accuracy and transferability are limited. 
At the other extreme, 
methods based on first-principle quantum theory 
such as density functional theory (DFT)\cite{kohn1965ksdft} have been proposed, 
the most well-known example being the 
\textit{ab initio} molecular dynamics (AIMD)\cite{car1985cpmd} scheme. 
These methods promise to be much more accurate but they are also computationally expensive, 
limiting our ability to handling systems of hundreds to thousands of atoms only.
Until recently, 
the drastic disparity between these two approaches in terms of accuracy and computational cost 
has been a major dilemma to be confronted with in molecular simulation.

Recent advances in machine learning, particularly deep learning, 
have ushered some new hope in addressing this dilemma\cite{schutt2016dtnn,behler2007bpnn,faber2017fast,chmiela2017machine,handley2014review,behler2015review,rupp2012CM, ramakrishnan2015CM,huang2016BAML,hansen2015Bob,bartok2013SOAP,smith2017ani}.
Several promising new ideas have been suggested, in which deep neural networks are used to represent the potential energy surface.
Of particular interest are the Behler-Parrinello neural network (BPNN)\cite{behler2007bpnn} and the deep tensor neural network (DTNN)\cite{schutt2016dtnn}. 
BPNN uses the so-called symmetry functions as input and a standard neural network as the fitting function; 
DTNN, on the other hand, uses as input a vector of nuclear charges and an inter-atomic distance matrix, and introduces a sequence of interaction passes where 
``the atom representations influence each other in a pair-wise fashion"\cite{schutt2016dtnn}. 
Both methods are able to predict with chemical accuracy the potential energy surface of
materials in condensed phase, in the case of BPNN,
and of small organic molecules, in the case of DTNN. 
The construction of the local symmetry functions for BPNN contains 
an \textit{ad hoc} and often tedious component 
where hand-crafted fitting functions and human intervention are required. 

In this work, we develop a new method, called Deep Potential (DP), 
that successfully addresses the inadequacies of the existing models.
Deep Potential is a simple, yet general, 
end-to-end deep neural network representation of a many-atom potential energy surface. 
The network uses as input  the raw coordinates  of the atoms in a proper frame of reference, 
and naturally respects the symmetries of the system.  
Promising results are obtained in a variety of test cases, 
including small molecular isomers and condensed-phase systems. 
The Deep Potential method brings us closer to performing molecular modeling 
with the secure accuracy of first-principle based methods 
at a computational cost comparable to that of empirical potentials.

\section{\label{sec2}Results}
\begin{figure}
 \includegraphics[width=16cm]{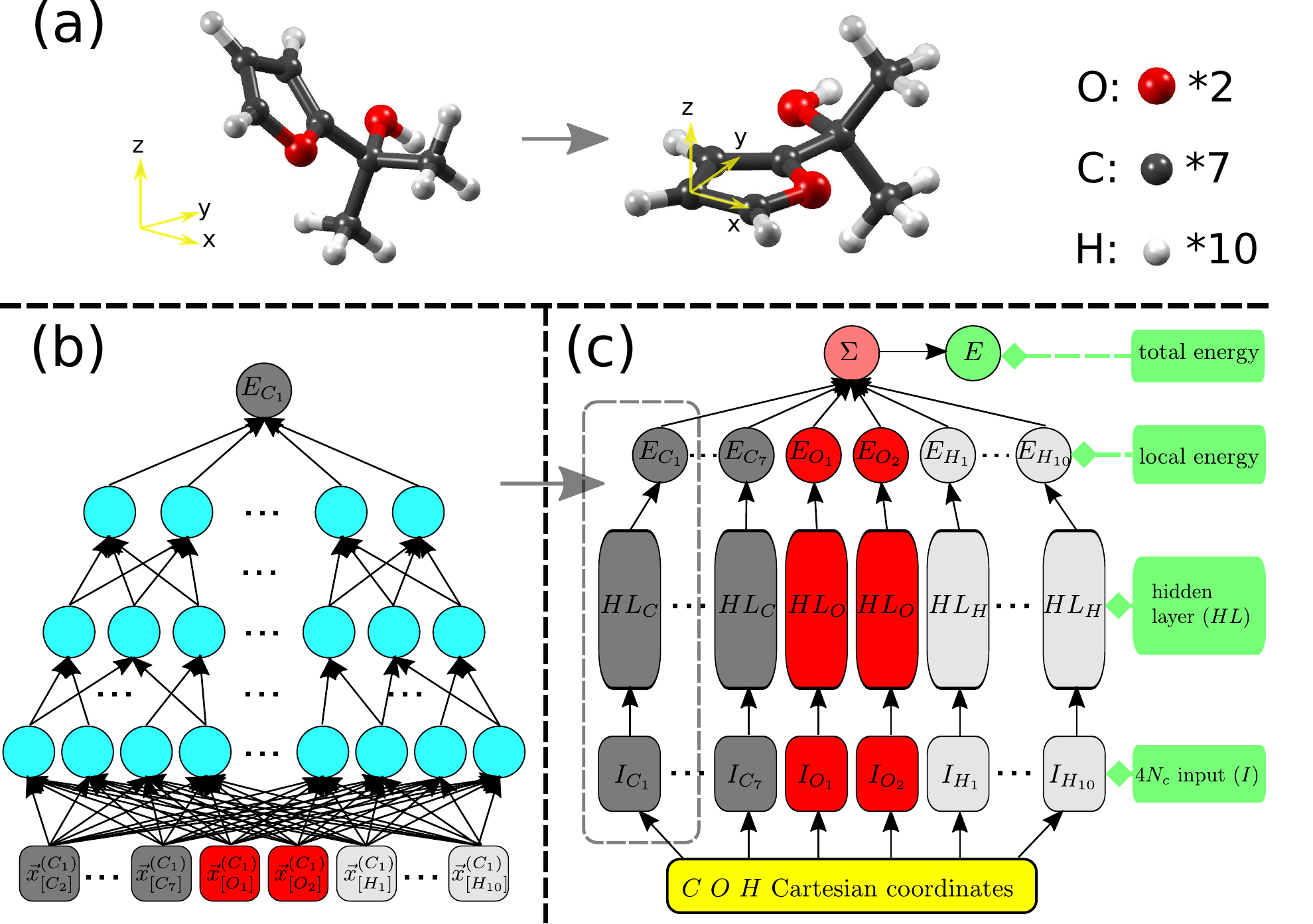}
 \caption{\label{fig1} Schematic diagram of Deep Potential, 
 using C${}_7$O${}_2$H${}_{10}$ as an illustrative example. 
 (a) Transformation from global to local reference frame for a Carbon atom; 
 (b) The sub-network structure for atom C${}_1$. 
 We use the notion $\vec{\bm{x}}^{(C_1)}_{[A_i]}$ to represent the coordinates of 
 the atom $A_i$ in the list of neighbors of atom C${}_1$. 
 The atoms A${}_i$ are assigned to different groups corresponding to 
 the different atomic species. 
 Within each one of these groups the atoms are listed in order of 
 increasing distance from C${}_1$; 
 (c) the full structure of the Deep Potential network.}
\end{figure}

{\bf Deep Potential framework.} Our goal is to formulate a general and direct end-to-end representation of the potential energy surface 
that uses the atomic configurations directly as the only input data.
The main challenge to achieve this goal is to design a deep neural network 
that obeys the important symmetries of the system, like it was achieved, 
for instance, with the convolutional neural network in pattern recognition problems \cite{krizhevsky2012CNN}. 
Besides the usual translational and rotational symmetries, we also have the permutational symmetry. 
We represent the potential energy surface by following the steps that are schematically indicated in \textbf{Fig. 1}.

For a system of $N$ atoms, our neural network consists of $N$ small, almost independent, copies. 
Each copy is a sub-network corresponding to a different atom in the system.
The size of the input data for a sub-network is at most $4N_c$, 
where $N_c$ is the number of atoms within the adopted cut-off radius $R_c$ (\textbf{Fig. 1 (b) and (c)}). 
If the number of atoms within $R_c$ fluctuates, $N_c$ is the largest fluctuating number.
We find that adopting a finite $R_c$ is sufficient in all the extended material systems considered here. 

Next, we introduce a local Cartesian coordinate frame for each atom.
In this frame, the atom under consideration is taken to be at the origin and is labelled as ``0". 
We fix the $x$ and $y$ axes in terms of the atom 0 and its two non-collinear nearest neighbors
that we label ``1'' and ``2'', respectively,
in order of increasing distance. 
For organic molecules, we exclude hydrogen atoms in the definition of ``1'' and ``2''.
The $x$-axis is defined by the 0-1 direction, 
the $z$-axis is along the direction of the cross product of 0-1 and 0-2,
and the $y$-axis is the cross product of $z$ and $x$. 
In this Cartesian frame, 
the coordinates of all the atoms falling inside the cut-off radius centered at the origin, 
excluding atom 0,
define the input data for the corresponding sub-network. 
In practice, we found that the combination $\{1/r,\cos\theta,\cos\phi,\sin\phi\}$, 
where $(r, \theta, \phi)$ are the polar coordinates, 
is a much better representation than the Cartesian coordinates $(x, y, z)$. 
This is so because $1/r$ automatically differentiates the atoms according to their inverse distances from the tagged atom at the origin. 
In some cases we find that it is sufficient to use the radial and angular coordinates 
of a smaller subset of atoms closer to the origin 
while keeping only the radial coordinates of all the other atoms within the cut-off radius.

The sub-networks are only coupled through summation in the last step of the scheme, when we compute the total energy.  
From a qualitative point of view, one can think about the sub-networks as providing different local energy contributions to the potential energy surface.
To preserve the permutational symmetry of the input, 
in each sub-network the atoms are first assigned to different groups 
corresponding to the different atomic species,  
and then within each one of these groups the atoms are sorted 
in order of increasing distance to the origin. 
Global permutational symmetry is preserved by assigning the same parameters to all the sub-networks corresponding to atoms of the same species.

The rest of the formulation of the deep neural network as well as the training procedure is fairly standard. 
In this work, we use a fully-connected feedforward neural network\cite{goodfellow2016dl}, 
sometimes combined with Batch Normalization \cite{ioffe2015BN}, 
in the architecture of the sub-networks. 
To train the network we employ stochastic gradient descent with the Adam optimizer\cite{Kingma2015adam}. 
We find that using hidden layers with the number of their nodes in decreasing order gives better results. 
Such phenomenon is similar to the coarse-graining process in convolutional neural network \cite{krizhevsky2012CNN}.
The compactness of the input data has the effect of reducing the number of parameters and thereby the complexity of the training process. 
For the systems considered here, 
it generally takes only a few hours on a NERSC Cori CPU node to train 
a small neural network capable of predicting the energy within chemical accuracy. 
See {\bf{Tab. 2}} for the architecture of the sub-networks and details of the training process.

{\bf Molecular system.} As a representative molecular system, we consider C${}_7$O${}_2$H${}_{10}$. 
Our goal is to predict the energies of all the isomers visited in a set of MD trajectories for this system, 
which comprises the largest ensemble of stable isomers in the QM9 database\cite{ruddigkeit2012qm9,ramakrishnan2014QM9}. 
The same collection of isomers was also used to benchmark the DTNN scheme. 
Following the convention of the DTNN benchmark, we measure the accuracy of the energy predictions 
using Deep Potential in terms of the mean absolute error (MAE) ({\bf{Tab. 1}}).

{\bf Condensed-phase systems.} 
To test the performance of Deep Potential in condensed phase,
we use data from EAM and AIMD simulations. 
In particular, we use 80000 MD snapshots in a 256 Cu atoms trajectory generated with the EAM potential. 
Consecutive snapshots are separated by 0.1 ps and we consider a trajectory 
in which the temperature is increased by 100 K every 200 ps under zero pressure. 
After reaching 2400 K, the temperature is decreased with the same schedule until the system reaches again a temperature of 500 K. 
We also consider 14000 snapshots along several AIMD trajectories for a solid Zr sample
in which the atoms are randomly displaced to model radiation damage. 
Finally, we use 100000 snapshots of a path-integral (PI) AIMD trajectory modeling 
liquid water at room temperature (300 K) and standard pressure (1 atm). 
In this trajectory, almost all the time steps are used for the training snapshots, so that a significant amount of correlation is present in the data.
For the condensed-phase systems, 
we follow the convention of BPNN and measure the accuracy of the energy predictions
with Deep Potential in terms of the root mean square error (RMSE) ({\bf{Tab. 1}}).

\section{\label{sec3}Discussion}
{\bf Generality.} By design Deep Potential is a very general framework.  
We obtain satisfactory results for finite and extended systems,
using data generated either by empirical potentials or by DFT simulations. 
Based on the experience gained from the test cases, we expect that the method should 
also work well for more complex systems, such as biological molecules. 
In addition, we could use for training more accurate data than DFT,
if available, 
such as, $e.g.$, data from Coupled-Cluster calculations with Single and Double and Perturbative Triple excitations\cite{Watts1993CCSDT} (CCSD(T)), or Quantum Monte Carlo simulations \cite{ceperley1986qmc}. 
\begin{table}[!ht]
\centering
\caption{Comparison of energy prediction obtained by Deep Potential and DTNN/BPNN}
\label{my-label}
\vspace{5pt}
\begin{tabular}{@{}c|c@{}}
\toprule
System                                 & MAE of DP (DTNN) (eV)\\ \midrule
C${}_7$O${}_2$H${}_{10}$ & 0.04 (0.07)\\
\toprule
System                                 & RMSE of DP (BPNN${}^*$) (meV/atom${}^{**}$) \\ \midrule
Cu                                        &  0.20 (0.20)          \\
Zr                                         &   0.09 (0.21)         \\
$\rm{H}_2\rm{O}$                &     1.8 (2)          \\ \bottomrule
\end{tabular}
\\
\vspace{5pt}
${}^*$ The results for Cu and Zr come from our implementation of BPNN with the symmetry functions for Cu in Ref. \cite{artrith2012Cu}; the result for $\rm{H}_2\rm{O}$ is from Ref. \cite{morawietz2016vdw}, which is tested on a different set of configurations. ${}^{**}$ The unit of RMSE for $\rm{H}_2\rm{O}$ is meV/molecule
\end{table}

{\bf Accuracy.} 
In all the test cases, the Deep Potential predictions reproduce the original data within chemical accuracy  ($\sim$ 1kcal/mol, or 0.04 eV) 
and compare well with either DTNN or BPNN ({\bf{Tab. 1}}). 
For C${}_7$O${}_2$H${}_{10}$, we obtain the up-to-date best result (0.04 eV) compared with the existing benchmark (0.07 eV). 
In the case of EAM Cu, we only use radial information to train the network, obtaining an accuracy comparable to BPNN. 
In the case of Zr with DFT data, Deep Potential with only radial information gives a better result (0.09 meV/atom) than BPNN (0.21 meV/atom), 
with the caveat that the symmetry functions we use for Zr are the same of those of Cu, 
$i.e.$, they are not redesigned for Zr. 
In the case of liquid water, including the radial and the first-shell angular information gives an error (1.8 meV/$\rm{H}_2\rm{O}$) that 
is one fourth of the error obtained with distance information only.

In our scheme, rotational and permutational symmetries are inposed by fixing operations.
The importance of the symmetry constraints can be assessed by selectively removing them.
For instance, removing rotational symmetry fixing in C${}_7$O${}_2$H${}_{10}$ gives an MAE of 0.05 eV, 
slightly larger than the optimal value for this molecule (0.04 eV).
On the other hand, removing permutational symmetry fixing in bulk Cu 
increases substantially the RMSE, from 0.2 meV/atom to 15.6 meV/atom.

{\bf Computational cost}. 
For a system with $N$ atoms and at most $N_c$ atoms inside $R_c$, 
the computational cost of Deep Potential is linear with $N$,
similar to EAM,
but requires a larger number of size independent operations.
In all systems, $N_c$ is size independent.
Using Cu as an example, both EAM and Deep Potential require $O(N)N_c$ operations to locate the atoms inside $R_c$. 
Within EAM, $\sim10NN_c$ multiplications are needed to fit the potential. 
Within Deep Potential, $\sim NN_c\log{N_c}$ operations are needed for the sorting, 
and $\sim100NN_c$ multiplications are needed for the calculation of the energy by the neural network. 
Both EAM and Deep Potential are highly parallelizable. 
Thus, large-scale MD simulations with Deep Potential are computationally feasible.

Within our framework we have the freedom to tune the size of the
sub-networks in order to achieve a good balance of computational cost
and accuracy of the predictions.
In the examples that we report, we use the sub-network sizes of {\bf Tab. 2},
but we could use significantly smaller sub-network sizes 
with a minor reduction of accuracy.
For instance, using a sub-network with 80-40-20-10 nodes 
in the hidden layers for C${}_7$O$_2$H$_{10}$,
we obtain a still acceptable MAE of 0.07 eV, 
to be compared with the MAE of 0.04 eV with the considerably
larger 600-400-200-100-80-40-20 sub-network.
The smaller sub-network reduces the computational cost by almost two orders of magnitude.

{\bf Scalability}. 
The scalability (and extensivity) of the energy of Deep Potential
is intimately related to the fact that the ``local energies'' $E_i$ depend only on a finite environment of site $i$.
This is consistent with physical intuition because
most of the interactions can be captured with a large enough $R_c$.
Typically including the first two shells of neighbors would allow us to describe covalent bonding interactions, such as bond stretching and bending, 
and dihedral angle forces. 
Van der Waals effects have longer range but we find that including
up to the third shell of neighbors is usually sufficient.
A small number of neighboring shells is also sufficient in metals 
due to screening effects.
Longer range effects are present in ionic and/or dipolar systems. 
The long-range part of the Coulombic interactions in such systems
is typically treated exactly with techniques such as the Ewald summation \cite{ewald1921ews}.
We have not included explicitly these effects in the current implementation
of Deep Potential, although in the cases in which these effects are important
they are included in the training data.
A possible way of explicitly including these effects in neural network potentials was presented in Ref. \cite{artrith2011zno}.
We leave the issue of how to include long-range effects into Deep Potential to future studies.
As a limited test of scalability, 
we use the Deep Potential model trained from the sample of 256 Cu atoms, mentioned earlier, to predict the potential energy of systems with
864, 2048, and 4000 atoms in periodic simulation cells. 
Comparison with EAM calculations for these systems gives an RMSE per atom of
0.10, 0.07 and 0.06 meV/atom, respectively. 

Deep Potential is  a simple, yet general, end-to-end deep neural network representation of
the potential energy function for atomic and molecular systems.  
It should enable us to evaluate the many-atom potential energy surface with the accuracy of quantum mechanics models at a computational cost comparable to that of empirical models.

\section{Methods}
{\bf Reference data sets.}
The AIMD data set and the training/testing protocals for C${}_7$O${}_2$H${}_{10}$
are available at 
\href{http://quantum-machine.org/datasets/}{http://quantum-machine.org/datasets/}. 
The simulation of Cu uses the MD package LAMMPS 2016 \cite{Plimpton1995lammps}. 
We adopt the EAM potential for Cu \cite{foiles1986eam} with an MD integration time step of 1 fs. 
The training system contains 256 atoms in a cubic cell with periodic boundary conditions.
We adopt a Nos\'e-Hoover thermostat and barostat following Ref. \cite{Tuckerman2006NPT}.
80000 configurations at intervals of 0.1 ps along the EAM trajectory for Cu are stored.
Of these, 72000 are used for training and 8000 to test the predictions.
The AIMD data for Zr are generated with VASP \cite{KG1996VASP} and consist of 14000 snapshots from 30 short trajectories that are used
to determine the threshold displacement energies (TDE) along symmetric directions. 
These simulations use Large supercells with 8000 atoms. 
Each trajectory runs for 1 ps with a variable time step ensuring that the atom with the largest velocity moves by less than 0.1 \AA~in each time step.
90\% of the corresponding snapshots are used for training and 10\% to test  the energy predictions.
The PI-AIMD trajectory for liquid water 
is obtained with Quantum Espresso \cite{QE-2009} interfaced with i-PI \cite{Ceriotti2014iPI}. 
The system contains 64 water molecules in a cubic box under periodic boundary conditions. 
The exchange-correlation functional PBE0 \cite{Carlo1999PBE0} is adopted, 
and the long-range dispersion interactions are approximated self-consistently with the Tkatchenko-Scheffler model \cite{TS2009TS}.
The generalized Langevin equation \cite{Ceriotti2011GLE}
in i-PI requires 8 beads for a converged representation of the Feynman paths. 
The trajectory is approximately 10 ps long with a time step of 0.48 fs. 
100000 snapshots are randomly selected along the PI-AIMD trajectory. 
Of these, 90000 are used for training and 10000 to test the energy predictions.

{\bf Details on the Deep Potential method.} 
After generating and renormalizing the input data from the atomic configurations, 
the implementation of the mothod follows standard deep learning procedures \cite{goodfellow2016dl}.
We use a fully connected neural network with
the activation function given by the rectified linear unit (ReLU).
We train the neural network with the Adam optimizer \cite{Kingma2015adam} with a batch size of 128. 
See {\bf Tab. 2} for the detailed architecture of the sub-networks, the total training epochs, 
the learning rate scheme, and the decay rate of the moving-average parameter for the Batch Normalization procedure. 
It is well known that Batch Normalization can effectively reduce the training time and improve the predictions of a deep and large neural network, 
like in the case of the C${}_7$O${}_2$H${}_{10}$ isomers, 
but for neural networks that are not deep and large, Batch Normalization is less effective. 
Thus, we use it for C${}_7$O${}_2$H${}_{10}$ and $\rm{H}_2\rm{O}$, 
but, after testing it, we decide not to use it for Cu or Zr. 
The adopted cutoff radii are 6.0~\AA, 7.0~\AA, and 5.8~\AA~ for Cu, Zr, and $\rm{H}_2\rm{O}$, respectively.

\begin{table}[!ht]
\centering
\caption{Details of training process}
\vspace{5pt}
\begin{tabular}{@{}c|c|c|c|c@{}}
\toprule
System                                 & architecture${}^*$   &  training epochs      &  LR scheme${}^{**}$ & Batch Normalization${}^{***}$ \\ 
C${}_7$O${}_2$H${}_{10}$ & 600-400-200-100-80-40-20  & 300  & (0.01, 0.96, 1.5) & used \\
Cu                                        & 80-40-20-10  &  600   &  (0.005, 0.96, 3.6)   & not used\\
Zr                                         & 80-20-5-5   & 1200 &  (0.005, 0.96,7.2) & not used \\
$\rm{H}_2\rm{O}$                &  160-40-10-10  & 300 & (0.01, 0.96, 1.6)  & used \\ \bottomrule
\end{tabular}
\\
\vspace{5pt}
${}^*$ The architecture is given by the number of nodes in each hidden layer. ${}^{**}$ The parameters ($a$, $b$, $c$) in the learning rate (LR) scheme are the starting learning rate, the decay rate, and the decay epoch, respectively. If the current training epoch is $x$, then the learning rate will be $a*b^{x/c}$. ${}^{***}$If Batch Normalization is used, the parameter for moving average will be $a*b^{0.6x/c}$
\end{table}

{\bf Acknowledgement}.  We thank Han Wang for providing the trajectory data of Zr and helpful discussions. The work of Han and E is supported in part by Major Program of NNSFC under grant 91130005, ONR grant N00014-13-1-0338, DOE grants DE-SC0008626 and DE-SC0009248. The work of Car is supported in part by DOE-SciDAC grant DE-SC0008626.




\begin{thebibliography}{10}

\bibitem{jones1924LJ}
J.~E. Jones, ``On the determination of molecular fields,'' in {\em Proceedings
  of the Royal Society of London A: Mathematical, Physical and Engineering
  Sciences}, vol.~106, pp.~463--477, The Royal Society, 1924.

\bibitem{stillinger1985SW}
F.~H. Stillinger and T.~A. Weber, ``Computer simulation of local order in
  condensed phases of silicon,'' {\em Physical Review B}, vol.~31, no.~8,
  p.~5262, 1985.

\bibitem{daw1984EAM}
M.~S. Daw and M.~I. Baskes, ``Embedded-atom method: Derivation and application
  to impurities, surfaces, and other defects in metals,'' {\em Physical Review
  B}, vol.~29, no.~12, p.~6443, 1984.

\bibitem{mackerell1998charmm}
A.~D. MacKerell~Jr, D.~Bashford, M.~Bellott, R.~L. Dunbrack~Jr, J.~D. Evanseck,
  M.~J. Field, S.~Fischer, J.~Gao, H.~Guo, S.~Ha, {\em et~al.}, ``All-atom
  empirical potential for molecular modeling and dynamics studies of
  proteins,'' {\em The Journal of Physical Chemistry B}, vol.~102, no.~18,
  pp.~3586--3616, 1998.

\bibitem{wang2000amber}
J.~Wang, P.~Cieplak, and P.~A. Kollman, ``How well does a restrained
  electrostatic potential (resp) model perform in calculating conformational
  energies of organic and biological molecules?,'' {\em Journal of
  Computational Chemistry}, vol.~21, no.~12, pp.~1049--1074, 2000.

\bibitem{van2001rff}
A.~C. Van~Duin, S.~Dasgupta, F.~Lorant, and W.~A. Goddard, ``Reaxff: a reactive
  force field for hydrocarbons,'' {\em The Journal of Physical Chemistry A},
  vol.~105, no.~41, pp.~9396--9409, 2001.

\bibitem{kohn1965ksdft}
W.~Kohn and L.~J. Sham, ``Self-consistent equations including exchange and
  correlation effects,'' {\em Physical Review}, vol.~140, no.~4A, p.~A1133,
  1965.

\bibitem{car1985cpmd}
R.~Car and M.~Parrinello, ``Unified approach for molecular dynamics and
  density-functional theory,'' {\em Physical Review Letters}, vol.~55, no.~22,
  p.~2471, 1985.

\bibitem{schutt2016dtnn}
K.~T. Schütt, F.~Arbabzadah, S.~Chmiela, K.~R. Müller, and A.~Tkatchenko,
  ``Quantum-chemical insights from deep tensor neural networks,'' {\em Nature
  Communications}, vol.~8, p.~13890, 2017.

\bibitem{behler2007bpnn}
J.~Behler and M.~Parrinello, ``Generalized neural-network representation of
  high-dimensional potential-energy surfaces,'' {\em Physical Review Letters},
  vol.~98, no.~14, p.~146401, 2007.

\bibitem{faber2017fast}
F.~A. Faber, L.~Hutchison, B.~Huang, J.~Gilmer, S.~S. Schoenholz, G.~E. Dahl,
  O.~Vinyals, S.~Kearnes, P.~F. Riley, and O.~A. von Lilienfeld, ``Fast machine
  learning models of electronic and energetic properties consistently reach
  approximation errors better than dft accuracy,'' {\em arXiv preprint
  arXiv:1702.05532}, 2017.

\bibitem{chmiela2017machine}
S.~Chmiela, A.~Tkatchenko, H.~E. Sauceda, I.~Poltavsky, K.~T. Sch{\"u}tt, and
  K.-R. M{\"u}ller, ``Machine learning of accurate energy-conserving molecular
  force fields,'' {\em Science Advances}, vol.~3, no.~5, p.~e1603015, 2017.

\bibitem{handley2014review}
C.~M. Handley, J.~Behler, {\em et~al.}, ``Next generation of interatomic
  potentials for condensed systems,'' {\em European Physical Journal B},
  vol.~87, p.~152, 2014.

\bibitem{behler2015review}
J.~Behler, ``Constructing high-dimensional neural network potentials: A
  tutorial review,'' {\em International Journal of Quantum Chemistry},
  vol.~115, no.~16, pp.~1032--1050, 2015.

\bibitem{rupp2012CM}
M.~Rupp, A.~Tkatchenko, K.-R. M{\"u}ller, and O.~A. Von~Lilienfeld, ``Fast and
  accurate modeling of molecular atomization energies with machine learning,''
  {\em Physical Review Letters}, vol.~108, no.~5, p.~058301, 2012.

\bibitem{ramakrishnan2015CM}
R.~Ramakrishnan and O.~A. von Lilienfeld, ``Many molecular properties from one
  kernel in chemical space,'' {\em CHIMIA International Journal for Chemistry},
  vol.~69, no.~4, pp.~182--186, 2015.

\bibitem{huang2016BAML}
B.~Huang and O.~A. von Lilienfeld, ``Communication: Understanding molecular
  representations in machine learning: The role of uniqueness and target
  similarity,'' {\em The Journal of Chemical Physics}, vol.~145, no.~16,
  p.~161102, 2016.

\bibitem{hansen2015Bob}
K.~Hansen, F.~Biegler, R.~Ramakrishnan, W.~Pronobis, O.~A. Von~Lilienfeld,
  K.-R. M{\"u}ller, and A.~Tkatchenko, ``Machine learning predictions of
  molecular properties: Accurate many-body potentials and nonlocality in
  chemical space,'' {\em The Journal of Physical Chemistry Letters}, vol.~6,
  no.~12, p.~2326, 2015.

\bibitem{bartok2013SOAP}
A.~P. Bart{\'o}k, R.~Kondor, and G.~Cs{\'a}nyi, ``On representing chemical
  environments,'' {\em Physical Review B}, vol.~87, no.~18, p.~184115, 2013.

\bibitem{smith2017ani}
J.~S. Smith, O.~Isayev, and A.~E. Roitberg, ``Ani-1: an extensible neural
  network potential with dft accuracy at force field computational cost,'' {\em
  Chemical Science}, vol.~8, no.~4, pp.~3192--3203, 2017.

\bibitem{krizhevsky2012CNN}
A.~Krizhevsky, I.~Sutskever, and G.~E. Hinton, ``Imagenet classification with
  deep convolutional neural networks,'' in {\em Advances in Neural Information
  Processing Systems}, pp.~1097--1105, 2012.

\bibitem{goodfellow2016dl}
I.~Goodfellow, Y.~Bengio, and A.~Courville, {\em Deep learning}.
\newblock MIT Press, 2016.

\bibitem{ioffe2015BN}
S.~Ioffe and C.~Szegedy, ``Batch normalization: accelerating deep network
  training by reducing internal covariate shift,'' in {\em Proceedings of The
  32nd International Conference on Machine Learning (ICML)}, 2015.

\bibitem{Kingma2015adam}
D.~Kingma and J.~Ba, ``Adam: a method for stochastic optimization,'' in {\em
  Proceedings of the International Conference on Learning Representations
  (ICLR)}, 2015.

\bibitem{ruddigkeit2012qm9}
L.~Ruddigkeit, R.~Van~Deursen, L.~C. Blum, and J.-L. Reymond, ``Enumeration of
  166 billion organic small molecules in the chemical universe database
  gdb-17,'' {\em Journal of Chemical Information and Modeling}, vol.~52,
  no.~11, pp.~2864--2875, 2012.

\bibitem{ramakrishnan2014QM9}
R.~Ramakrishnan, P.~O. Dral, M.~Rupp, and O.~A. Von~Lilienfeld, ``Quantum
  chemistry structures and properties of 134 kilo molecules,'' {\em Scientific
  Data}, vol.~1, 2014.

\bibitem{Watts1993CCSDT}
J.~D. Watts, J.~Gauss, and R.~J. Bartlett, ``Coupled‐cluster methods with
  noniterative triple excitations for restricted open‐shell hartree–fock
  and other general single determinant reference functions. energies and
  analytical gradients,'' {\em The Journal of Chemical Physics}, vol.~98,
  no.~11, pp.~8718--8733, 1993.

\bibitem{ceperley1986qmc}
D.~Ceperley and B.~Alder, ``Quantum monte carlo,'' {\em Science}, vol.~231,
  1986.

\bibitem{artrith2012Cu}
N.~Artrith and J.~Behler, ``High-dimensional neural network potentials for
  metal surfaces: A prototype study for copper,'' {\em Physical Review B},
  vol.~85, no.~4, p.~045439, 2012.

\bibitem{morawietz2016vdw}
T.~Morawietz, A.~Singraber, C.~Dellago, and J.~Behler, ``How van der waals
  interactions determine the unique properties of water,'' {\em Proceedings of
  the National Academy of Sciences}, p.~201602375, 2016.

\bibitem{ewald1921ews}
P.~P. Ewald, ``Die berechnung optischer und elektrostatischer
  gitterpotentiale,'' {\em Annalen Der Physik}, vol.~369, no.~3, pp.~253--287,
  1921.

\bibitem{artrith2011zno}
N.~Artrith, T.~Morawietz, and J.~Behler, ``High-dimensional neural-network
  potentials for multicomponent systems: Applications to zinc oxide,'' {\em
  Physical Review B}, vol.~83, no.~15, p.~153101, 2011.

\bibitem{Plimpton1995lammps}
S.~Plimpton, {\em Fast parallel algorithms for short-range molecular dynamics}.
\newblock Academic Press Professional, Inc., 1995.

\bibitem{foiles1986eam}
S.~Foiles, M.~Baskes, and M.~Daw, ``Embedded-atom-method functions for the fcc
  metals cu, ag, au, ni, pd, pt, and their alloys,'' {\em Physical Review B},
  vol.~33, no.~12, p.~7983, 1986.

\bibitem{Tuckerman2006NPT}
M.~E. Tuckerman, J.~Alejandre, R.~L\'opez-Rend\'on, A.~L. Jochim, and G.~J.
  Martyna, ``A liouville-operator derived measure-preserving integrator for
  molecular dynamics simulations in the isothermal–isobaric ensemble,'' {\em
  Journal of Physics A: Mathematical and General}, vol.~39, no.~19, p.~5629,
  2006.

\bibitem{KG1996VASP}
G.~Kresse and J.~Furthm\"uller, ``Efficient iterative schemes for ab initio
  total-energy calculations using a plane-wave basis set,'' {\em Physical
  Review B}, vol.~54, pp.~11169--11186, 1996.

\bibitem{QE-2009}
P.~Giannozzi, S.~Baroni, N.~Bonini, M.~Calandra, R.~Car, C.~Cavazzoni,
  D.~Ceresoli, G.~L. Chiarotti, M.~Cococcioni, I.~Dabo, A.~{Dal Corso},
  S.~de~Gironcoli, S.~Fabris, G.~Fratesi, R.~Gebauer, U.~Gerstmann,
  C.~Gougoussis, A.~Kokalj, M.~Lazzeri, L.~Martin-Samos, N.~Marzari, F.~Mauri,
  R.~Mazzarello, S.~Paolini, A.~Pasquarello, L.~Paulatto, C.~Sbraccia,
  S.~Scandolo, G.~Sclauzero, A.~P. Seitsonen, A.~Smogunov, P.~Umari, and R.~M.
  Wentzcovitch, ``Quantum espresso: a modular and open-source software project
  for quantum simulations of materials,'' {\em Journal of Physics: Condensed
  Matter}, vol.~21, no.~39, p.~395502 (19pp), 2009.

\bibitem{Ceriotti2014iPI}
M.~Ceriotti, J.~More, and D.~E. Manolopoulos, ``i-pi: A python interface for ab
  initio path integral molecular dynamics simulations,'' {\em Computer Physics
  Communications}, vol.~185, pp.~1019--1026, 2014.

\bibitem{Carlo1999PBE0}
C.~Adamo and V.~Barone, ``Toward reliable density functional methods without
  adjustable parameters: The pbe0 model,'' {\em The Journal of Chemical
  Physics}, vol.~110, no.~13, pp.~6158--6170, 1999.

\bibitem{TS2009TS}
A.~Tkatchenko and M.~Scheffler, ``Accurate molecular van der waals interactions
  from ground-state electron density and free-atom reference data,'' {\em
  Physical Review Letters}, vol.~102, p.~073005, 2009.

\bibitem{Ceriotti2011GLE}
M.~Ceriotti, D.~E. Manolopoulos, and M.~Parrinello, ``Accelerating the
  convergence of path integral dynamics with a generalized langevin equation,''
  {\em The Journal of Chemical Physics}, vol.~134, no.~8, p.~084104, 2011.

\end{thebibliography}

\end{document}